\documentclass[fleqn,12pt,twoside,epsfig]{article}
\usepackage{espcrc1}

\usepackage{graphicx}
\usepackage[figuresright]{rotating}

\newcommand{\AmS}{{\protect\the\textfont2
  A\kern-.1667em\lower.5ex\hbox{M}\kern-.125emS}}

\title{ THE QCD PHASE DIAGRAM, EQUATION OF STATE,\\ 
AND HEAVY ION COLLISIONS  }

\author{ E.Shuryak \\
               State University of New York,\\
                   Stony Brook NY 11790
        \thanks{This work is partially
supported by US DOE, by the grant No. DE-FG02-88ER40388.}
}

\begin{document}

\maketitle

\begin{abstract}
After some historic remarks and a brief summary of recent 
theoretical news about the
QCD phases, we turn to the issue of 
$freeze-out$ in heavy ion collisions. We argue that the chemical
freeze-out line should actually consists of two crossing lines
of different nature. We also consider some 
inelatic reactions which occure 
 $after$
chemical freeze-out, emphasizing the role of 
overpopulation of pions.
The $hydrodynamics$ (with or without hadronic
afterburner)  explaines SPS/RHIC data on radial and elliptic flow
in unexpected details,for different particles,
collision energies, and impact parameters.
 Apart of Equation of State (EoS),
it has basically no free parameters. The
 EoS which describe these data best agrees
quite well with the lattice predictions, with the QGP latent heat
$\Delta\epsilon\approx 800 \, Mev/fm^3$.
Other phenomena at RHIC, such as ``jet quenching'' and
huge ellipticity  at large $p_t$, also point toward very rapid
entropy production. Its mechanism remains an outstanding
open problem: at the end we
discuss  recent application of the instanton/sphaleron
 mechanism. 
The gg collisions with $\sqrt{s}=2-3 \, GeV$  may result not in
mini-jets but rather in production of sphaleron-like
gluomagnetic clusters, which 
are classically unstable and
 promptly decay into several gluons and quarks, in sperical 
mini-Bangs.
\end{abstract}

\section{Introduction: the beginning of our field}

  Since I happen to present the last talk at this conference, 
let me start with some historic remarks, on the early days
of our field in general and 
the role of Helmut Satz in particular.  I know Helmut longer than  
the topic of  conference, ``Statistical QCD'', existed: 
we first met I think in 1975 in Dubna, a year after my Ph.D.\footnote{
Very characteristically, I could not find any material from that
time,
by Helmut had made and kept very nice photos: thanks Helmut once
again.}.
 Already then I have been stronhly influenced
 by his general style, which I would
characterize as `` forceful but
kind''. (At this meeting he has demonstrated the same qualities
once again, as the chairman of the last session, 
saving myself from a cascade of difficult questions after this talk.)

  Speaking of our field, let me define it
as the intercept of (i)  
the theory of hot/dense matter $different$ from the usual
hadronic matter (such as nuclear matter); with (ii) 
the
phenomenology of high energy heavy ion collisions\footnote{There are
also phenomenology of neutron stars and cosmological applications: 
but they are still struggling to be in the mainstream.
}.
 Of course the first use of thermodynamics
for hadronic collisions were suggested 50 years ago by
 {\it Fermi} \cite{fermi}. Soon after
  {\it Landau} \cite{landau} argued that hydrodynamics
should also be applicable, provided the system 
is {\em macroscopically large}, its size $L$ is much larger than
a micro scale $l$, the mean free path.
(Below I will return to the fundamental question of our field,
whether the system created in heavy
ion collisions does or does not satisfy this criterium. )
 When
 the first hadronic resonance, $\Delta$, has just been discovered, {\it  Landau and
S.Z.Belenkij} \cite{landau} promtly introduced the ``resonance gas'',
supporteed by the Beth-Ulenbeck result for the second virial coefficient.
 {\it  Hagedorn} \cite{Hagedorn} in 1960's developed this idea into
 dual bootstrap approach for resonances, and
introduced his famous limited T. As it was understood later, it is indeed
 a maximal T in a {\em hadronic phase}
  in which there is confinement and QCD strings.
 It was very important that from the start Hagedorn and 
others made
   practical  applications of the statistical model
  to various hadronic reactions\footnote{I also studied some of them as part
of my Ph.D. in early 1970's,
 and concluded that for reactions like low energy $\bar p p\rightarrow 
n\pi$ 
the thermodynamics indeed works remarcably well, provided (i) one uses
microcanonical approach for strangeness etc; (ii) the freeze-out
happens at fixed desnity, not volume. Both of them have surfaced now again.}.

   Of course in the early 70's, when  QCD has been discovered,
its ideas were soon applied to hot/dense matter.
   Based on asymptotic freedom,
  { Collins and Perry} \cite{CP}  suggested that the  high T
and/or density matter should be close to
the ideal gas. The  first perturbative corrections
were evaluated \cite{firstpert}.
The part of gluonic  polarization tensor due to high T  \cite{Shu_zhetp}
(unlike that due to virtual gluonic loops)
 $screens$ the charge: thus the name of
this new phase,
{\em Quark-Gluon Plasma}, QGP. Its excitations are quasiparticles
similar to ``plasmons'' etc of the usual plasma.
Furthermore, it was found in  \cite{Shu_zhetp} that static
magnetic field is not screened: thus the infrared divergences and other
non-perturbative phenomena  survive in the
 magnetic sector. First resummations of (what was later called)
Hot Thermal Loops into a plasmon term have also been
done   \cite{Shu_zhetp,Kapusta} by the end of 1970's. (Its consistent
application to many other
problems  have been worked out by
 { Braaten, Pisarski} and others a decade later.)
My other paper \cite{Shu_qgp}
 addressed such practical questions   
as new flavor, photon and dilepton production from QGP, as well
as the first take on the $J/\psi$
suppression\footnote{ Of course, the famous 1986 paper by
Matsui and Satz have superseeded it, by 
showing that in sufficiently hot QGP the $J/\psi$
is not even bound. However later, Satz, Kharzeev and others
have also worked out gluonic  excitation to continuum.}, by gluon
excitation similar to photoeffect.

By the end of 70's Helmut was the first to realize that in order for this 
field to be  shaped 
one has to take it all together: he called the 
 first Quark Matter conference - the { famous  Bielefeld meeting} of
1980 (soon reinforced by the next one in 1982) - which have defined
the field and created our community.

Since at the time
 I still was, so to say, in a confined phase, 
I  could not attend these 
meetings.
As  I also felt  that something should be 
done at that moment, I wrote the
  first review article \cite{Shu_80} in which  available
theory results  were combined with potential applications to high
energy collisions.

\section{The  QCD phase diagram, version 2001}
 Let me start with general remarks about the phases of QCD. Basically,
starting from the hadronic phase
 there 
are three directions to go: (i) {\em High T}  leading 
to QGP, the perturbative phase without any condensates. (ii)
 {\em High density} direction,
 which leads us to very rich world dominated by
quark pairing providing a set of Color Superconducting phases,
with broken  color group, sometimes with broken chiral
symmetry or even broken
parity. 
(iiI) {\em Increasing number of light quarks $N_f$}
which  brings another
 strange  conformal world,  without condensates but with the infrared fixed
point. Very few studies have been done,
 on the lattice
\cite{iwasaki} and with instantons (see \cite{SS_98}): both have indicated that
the critical $N_f$ can be as low as 7 or 5.

All this large variety of phases (including of course the hadronic
phase
we live in) and their boundaries 
can  be understood with remarcably small set of dynamical tools.
Basically, if one is interested in asymptotically large
$T,\mu$, the one-gluon exchange should be used. If not, the {\em
instanton-induced} t'Hooft Lagrangian is enough. It generates
attraction in $three$ different channels: (i) {\em quark-antiquark}, leading
to chiral symmetry breaking; (ii) {\em quark-quark} channel leading to color
superconductivity; and (iii) attraction between {\em instanton and 
anti-instanton}, mediated by light quarks, which leads to their pairing 
in QGP and conformal phase. The competition of these three attractions
defines the phase boundaries.

Let me now make a quick update on recent news, starting with
the T direction. Important work by the Bielefeld
group
(see Laermann, this proceedings) have fixed the critical mass of the quark,
in the case when it is the same for 3 flavors. Its value indicate
that in QCD we would nearly certainly have a cross-over, not  the
first order transition.

Addressing the fate of this cross-over line  along 
the $\mu$ direction, Fodor and Katz \cite{FK} have found the 
  first numerical evidences that the location of the critical endpoint
E is at
$T\approx 160 \, MeV, \mu_B\approx 700 MeV$, see fig.\ref{chemfreeze}(a).

Going further to the high density  we find intense theoretical
 activity.
The main news is the ``stress'' which is provided by the non-zero
strange quark mass on the Color-Flavor-Locked (CFL) phase
leads to spontaneous breaking of the $P-parity$, with the non-zero
   {\em kaon condensate}, as first shown by
Bedaque and Schafer  \cite{BS}. More details on the resulting
phase diagram in the plane of chemical potentials for
baryon and electric charge
has been
provided by Kaplan ad Reddy \cite{KR}: the most surprising feature is
 that
the recovery of pure CFL phase happens only at extremely high density,
at $\mu_b$ of the order of millions of GeV. (One will find details on it
in T.Schafer's talk in this proceedings.)

  The second time in this field there were two papers from the Stony
Brook and the (now) MIT teams,
 submitted to hep-ph on the same day\footnote{The
first
time, in 1998, those were papers by Alford,Rajagopal and Wilczek, and
Rapp,Schafer, Shuryak and Velkovky, which have revived this field. }:  both
are the first attempts to ``try the ice'' by looking into
possible {\em crystalline phases of quark matter}. 
This time however the content
of the papers are rather different.
Rapp, Zahed and myself \cite{RSZ}
have looked at frozen $\bar q q$ (or sigma)
field with non-zero momentum: a phenomenon
similar to the so called Pierls instability in 1d and Overhauser spin
waves
in 2d. We found that given very strong pairing force, it may well
compete
with the 2-flavor Color Superconductivity at $\mu_b\sim 1200 \, MeV$.

Alford, Bowers and Rajagopal  \cite{ABR} have considered a different
crystal, in which $q q$ (or diquark) condensate has the
 non-zero momentum: it is similar to the so called LOFF phase
proposed (but not observed yet) for ordinary superconductors.
Its tentative place is a narrow region
along the border between 2- and 3-flavor-like
superconducting phases. 

There are multiple works on unusual properties of excitations. Let me
single
out two nice papers by Son, Stephanov and Zhitnitsky \cite{SSZ} who have
addressed
the $\eta'$ and instantons at very high $\mu$. They have shown this to
be
the first example of relatively simple ``instanton liquid'', 
in this case a 4-d
dilute
Coulomb plasma with the $\eta'$ being the exchange field which gets
massive due to the Debye mechanism. They have also shown
that in this phase there can exist metastable domain walls made of
the $\eta'$ field: maybe bubbles made of such membranes can
ocasionally be produced in heavy ion collisions.

\section{Heavy ion collisions}
\subsection{ Physics of the freeze-out}  
As shown in several talks at this conference, 
statistical description of particle composition 
which worked well before  also does so at
 RHIC. The only new number to know
is $\mu_b\approx 45 \, MeV$: the resulting predictions agree with measured
particle ratios quite well.

However, {\em Does the success of statistical description
 imply that we deal with  macrosopically large systems?} Not really: the
same thermal models
provide nearly equally good description
of  hadronic yields from very low energy reactions like $\bar p p$
annihilation at rest to high energy
 $e^+e^-$ annihilation into hadrons. In the latter case we are quite
sure
the system is rather dilute: and thus excellent predictions of the
thermal model here remains (a long standing) puzzle.   

\begin{figure}[!tbp]
   \begin{center}
      \includegraphics[height=2.in]{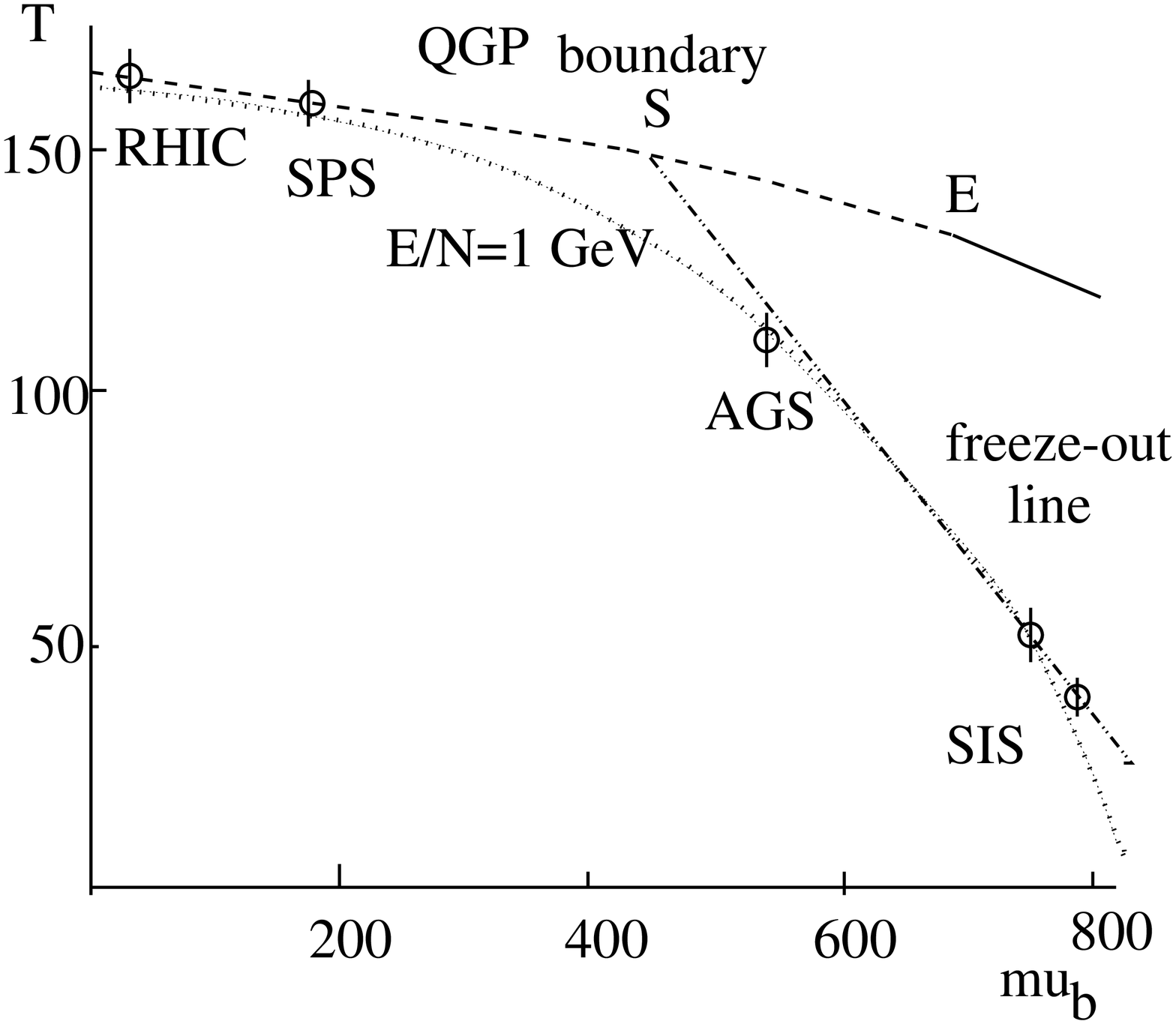}
      \includegraphics[height=2.in]{k+topi+.eps}
    \end{center}
      \caption{
      \label{chemfreeze}
    (a) A compillations of chemical freeze-out points on the phase
diagram temerature T (MeV) versus
the baryon chemical potential $\mu_b$ (MeV). The dotted line is the
Cleymans-Redlich ``semi-circle'' E/N= 1 GeV. The dashed and solid line
show crossover and 1-st order phase boundary, the dash-dotted
line is the kinetic chemical freezeout line in the hadronic phase. 
The crossing (marked S) is the ``strangest point'' corresponding
to the maximum in fig.(b). 
  (b) The compillation of data on $K^+/\pi^+$ ratio (P.B.Munzinger et
al, hep-ph/0106066), the curve corresponds to predictions
of the statistical model {\em on Cleymans-Redlich ``semi-circle''}.
      }
\end{figure}

Fitting the
matter composition from particular data set, one gets a point at the $T-\mu_b$
diagram. {\bf What kind of pattern one should eventually
see when many such points are collected?} Cleymans and Redlich
\cite{Redlich-Thermal}
have suggested a smooth
 chemical freeze-out line at $T-\mu_b$ plane:
they have empirically shown that it is nearly a {\it semi-circle}
approximately correspondnf to
 energy per particle E/N about 1 GeV. However  I would  suggest a 
 different picture (see Fig. \ref{chemfreeze}(a)), namely  
{\em two  crossing lines}, (i) the QGP phase boundary; and (ii)
chemical freeze-out line defined by hadronic rates. Their nature
is quite different: the first is a place of rapid changes in
thermodynamics,
the second is of kinetic nature and depends on things like expansion
rate, and is in principle  different for different nuclei and
impact
parameters. Although at the phase diagram it looks like two lines
are nearly parallel and their crossing would be difficult  to detect,
it looks more promising
 in other observables. Strangeness content is one of them:
see the $K^+/\pi^+$ ratio in Fig.\ref{chemfreeze}(b), in which the change of behavior
is really 
striking. My prediction is that instead of a round maximum shown
(corresponding 
to this cemi-circle), we should see instead a peak sue to the crossing of two
lines\footnote{In a particular model similar behavior but with $two$
discontinuities has been
demonstrated previously
by M.Gazdzicki et al, hep-ph/0006236. }.
Taking more data around the maximum,
which we may call ``the strangest point'' marked S  in Fig.\ref{chemfreeze}(a)
 is therefore very interesting
and directly linked to the QGP phase boundary.  

What happens {\em after chemical freeze-out} is still hotly debated. As
emphasized by Rafelski, in spite of baryon-rich environment
at SPS, the anti-hyperon production is not strongly suppressed
by annihilation, and their
spectra show the same shape as hyperon's, even a small $p_t$.
The same is true for $\bar p$ as well. Logically speaking, there
are two options: (i) either there is no interaction after
 chemical freeze-out  (as Rafelski  argued); or (ii) there exist
back reactions, compensating for the annihilated anti-baryons. 
Rapp and myself  as well as C.Greiner \cite{antibaryons}
have shown that (contrary to disbelief of many) the rates of
those are quite sufficient to nearly completely compensate
the annihilation. The key element here is the well {\em pion
overpopulation} phenomenon: after chemical equilibration point
(at which $\mu_\pi$ is set to zero), the number of pions is 
conserved while T goes down: it means that $\mu_\pi$
grows, reaching about 70 MeV. Since annihilation $\bar N N\rightarrow n\pi$ 
include $n\approx 6$ pions, the inverse rate gets a big help from
$exp(n\mu_\pi/T)$ factor.
   
 ``Life after chemical freeze-out'' has been recently studied 
by Derek Teaney \cite{Teaney-Chemical}, 
as part of hydro studies to be discussed below. 
Incorporating non-zero $\mu$'s for $all$ species, he calculated
 the thermodynamics of the resulting chemically non-equilibrium
resonance gas. (As above for pions, particle numbers per entropy
are kept the same as at the chemical freeze-out point.) It turns out
$p(\epsilon)$ (and thus hydro) hardly change, but $\epsilon(T)$
(and thus spectra) did
changed
quite a lot. Remarkably, new EoS of the resonance gas seem to match
that of RQMD: if it is used for description of the hadronic phase,
the results no longer depend on where transition from hydro to RQMD
takes place.

\subsection{ Heavy ion collisions: flows and EoS} 

 According to Landau, a
decisive test of the macroscopic  behavior is
 $hydrodynamical$ flows.
{\em Is this criterium satisfied by elementary pp or  $e^+e^-$ collisions?}
 With appearence of the first  multiparticle production data
from accelerators in 1970's,
the answer seemed affirmative at first \cite{ES_hydro},\cite{Heretic}, 
but more
accurate $\pi,K,p$
 spectra from
 ISR had  shown  \cite{ShurZh}  that in this case
there is no sign of transverse expansion\footnote{ It has been argued
 in
\cite{ShurZh}
that
the vacuum pressure may balance transverse pressure: the suggested picture 
resembled rather a string-like  model  than a hydro explosion.} of
matter. Beccatini at this conference have shown fits to more modern
high energy data, which indicate something like transverse flow, but with
 the  velocity  only $v_t\sim 0.2$ or so, 
much less than hydro predicts (and the RHIC data show),  $v_t\sim 0.6$.

Early  BEVALAC data have displayed collective phenomena
such as directed flow and ``squeeze-out'', and
 hydro description had been
attempted \cite{Greiner}.
 Eventually however, it
has been also concluded that at such energies ($E\sim 1 AGeV$)
 the NN mean free
path $l_{m.f.p.} \sim 1.5\, fm$ is not small enough
compared to nuclear size $L\sim 6 \, fm$ to justify it.
 Hadronic cascade models took the lead from that time on, and
(such as event generators RQMD  we use) have provided good
description of AGS and SPS data.
  At AGS and even more so at SPS, much stronger collective
expansion effects have been observed. At SPS  the
 mean transverse 
velocity $v_t$ reached by matter is about 1/2.
Although it has been successfully described 
hydrodynamically \cite{Hung},  these works did not shed much
light at the EoS of very dense matter becauseat AGS/SPS conditions most of the
acceleration happens at late times, in  hadronic matter driven by
relativistic pions. So, its
hydro description is  just ``dual'' to hadronic cascades. 

The decisive shift occured  when
$non-central$ collisions at high
 energies have been studied experimentally at SPS and RHIC, and theoretically, 
by our group
\cite{TLS}, as well as by U.Heinz and collaborators  \cite{Kolb}.
 Based on these works and  RHIC data, we can now see that
we are 
 finally approaching the macrosopic regime. {\em Finally, we got a Bang, not
fizzle.}

 I would concetrate
here on our  last paper from the set \cite{TLS},
in which systematic study 
of radial and 
elliptic flow has been made, with detailed predictions of their
energy, centrality and $p_t$ dependence as a function of EoS
(the $only$ input for hydro). Our 
Hydro-to-Hadrons (H2H) model uses hydro in QGP and ``mixed'' phases,
but switch to hadronic cascade (RQMD) in the hadronic phase.
Thus, in contrast to others, we have {\em differential freeze-out}
of different species, which is very important aspect of the calculation. 
Its comparison with large set of SPS and RHIC data looks to me
very convincing: the model describes radial and elliptic flows in
great details. 
Of course, only few  examples can be given here: I selected 2 cases in which
the model was not anticipated to work, and surprized us.

\begin{figure}[!tbp]
   \begin{center}
      \includegraphics[height=3.0in, width=3.0in]{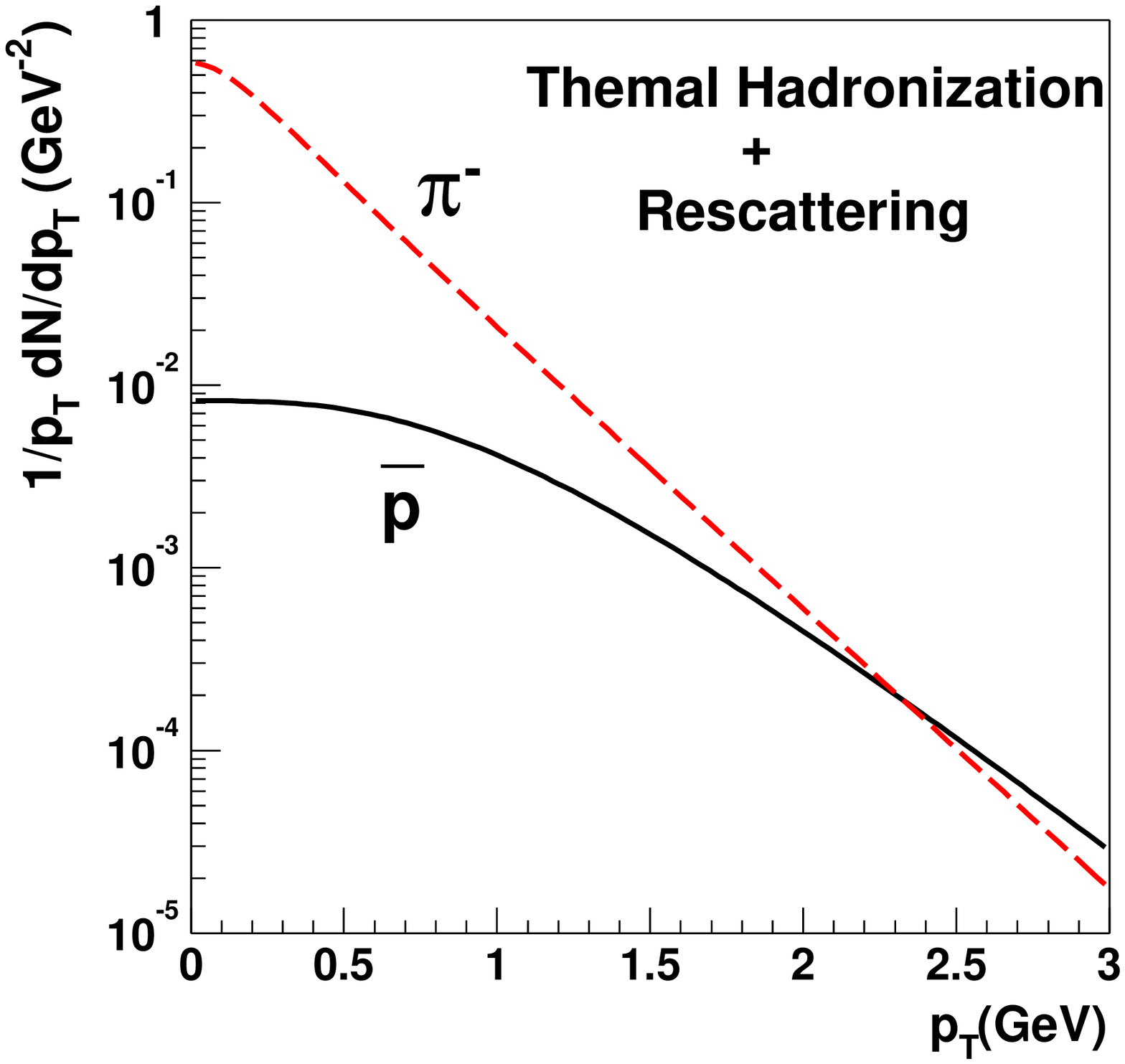}
      \includegraphics[height=3.0in, width=3.0in]{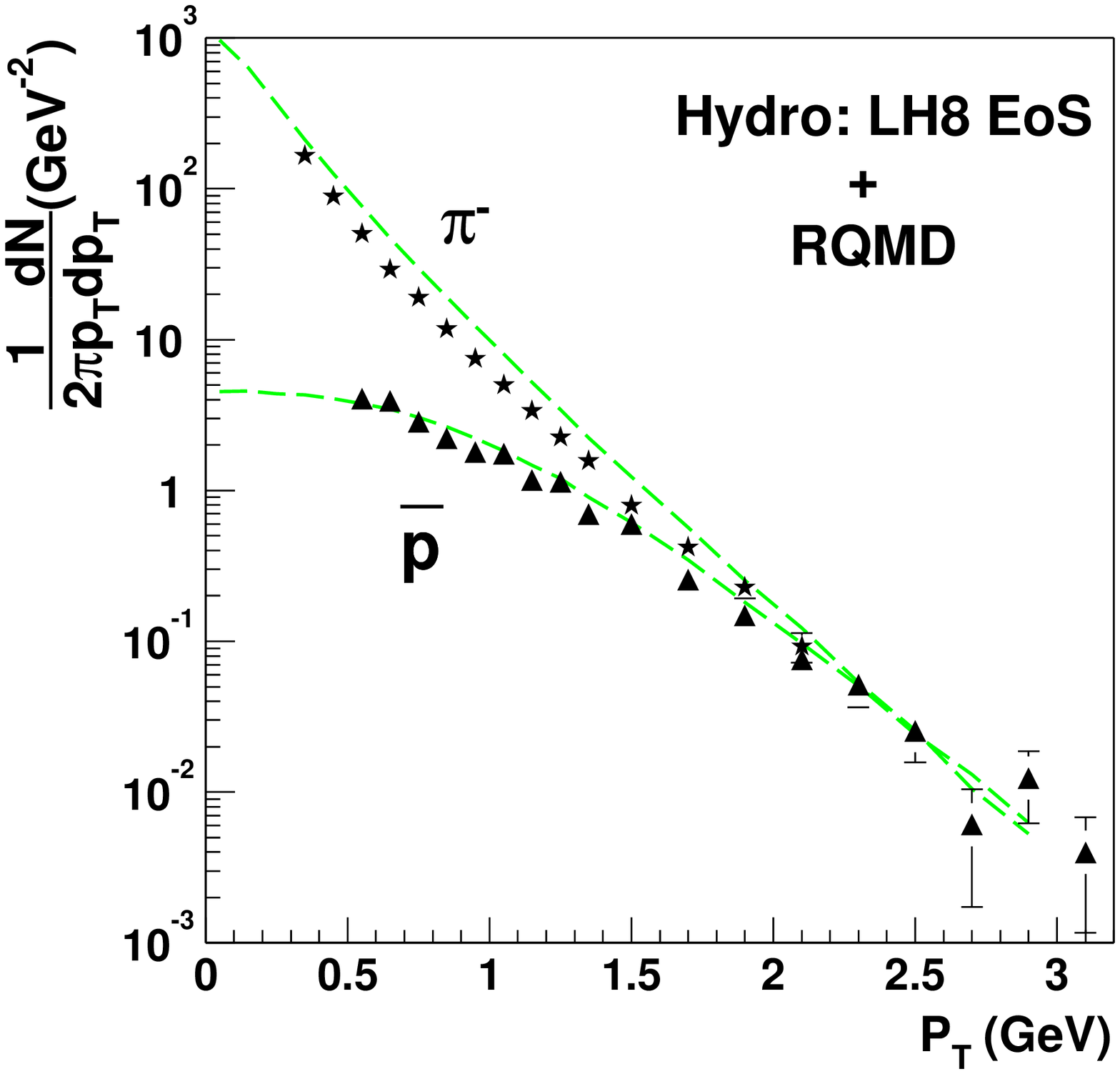}
    \end{center}
      \caption[A simple model of the $\pi^{-}$ and $\bar{p}$ 
       spectra]{
      \label{thermal-model}
      A comparison of $\pi^{-}$ and $\bar{p}$ spectra.
  (a) shows a simple thermal model with parameters discussed in 
      the text.
  (b) shows an absolutely normalized comparison of model
       and PHENIX spectra. 
      }
\end{figure}

One is ``crossing'' of $\pi^{-}$ and $\bar{p}$ spectra  
at $p_{T}\approx 2.0\,\mbox{GeV}$ shown in Fig.\ref{thermal-model}.
As shown in the l.h.s.,  even schematic hydro/thermal model with appropriate
transverse velocity $v_t\sim 0.6$ explains it. The r.h.s. shows
our predictions (both data and model have absolute normalizations: no
free
parameters). What is unusual about it is that hydro/thermal
description
seem to work even at the largest measured $p_t$: so far, no 
power-like tail due to hard processes is seen\footnote{One more argument that
it is not jet fragmentaion of any kind is this very point that they
would lead to
 the $\bar{p}/\pi^{-}$ 
ratio much less than 1.}.

\begin{figure}[h]
   \vskip 0.2in
   \includegraphics[width=2.6in, angle=0]{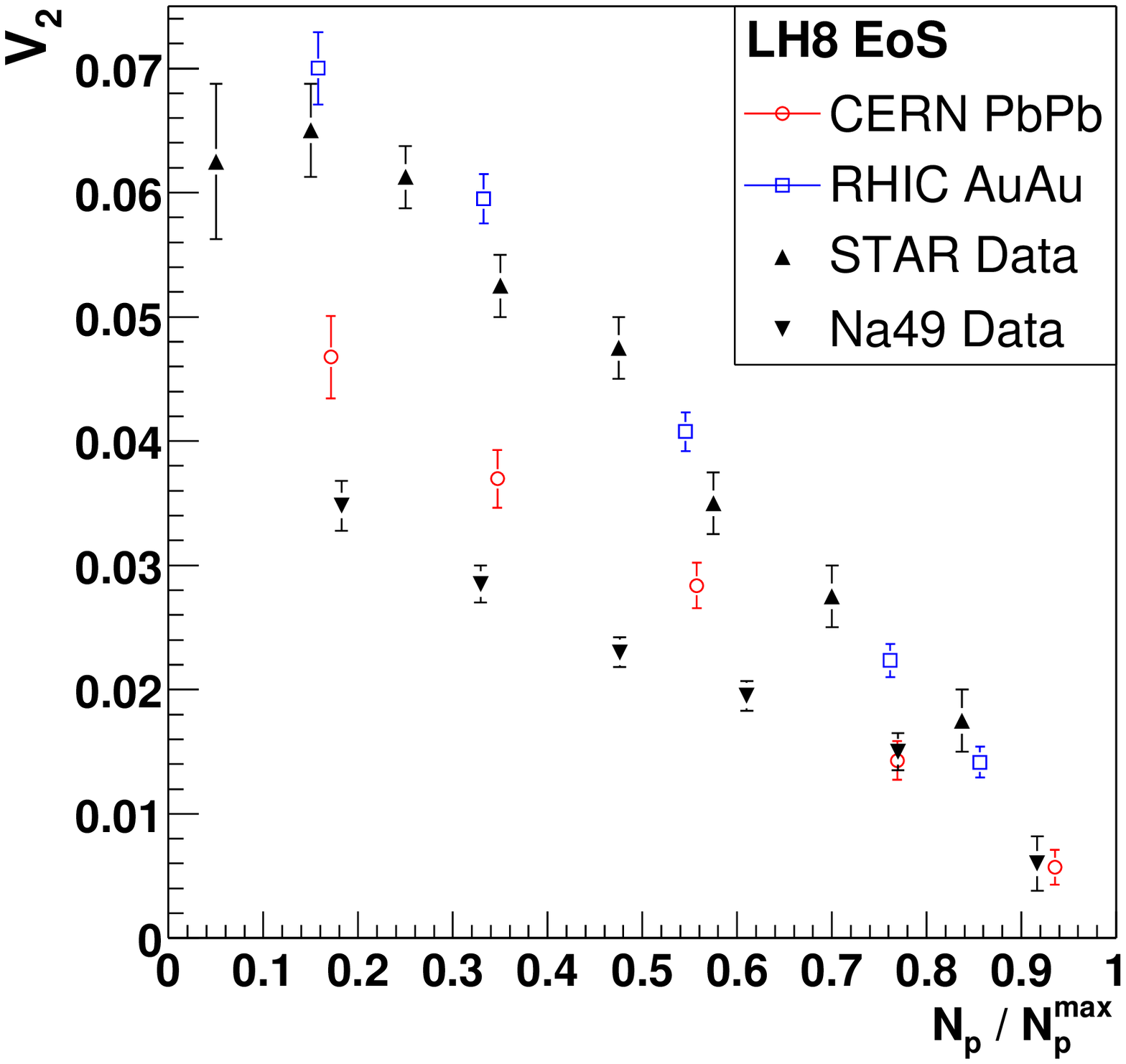}
      \includegraphics[height=2.6in,width=2.6in]{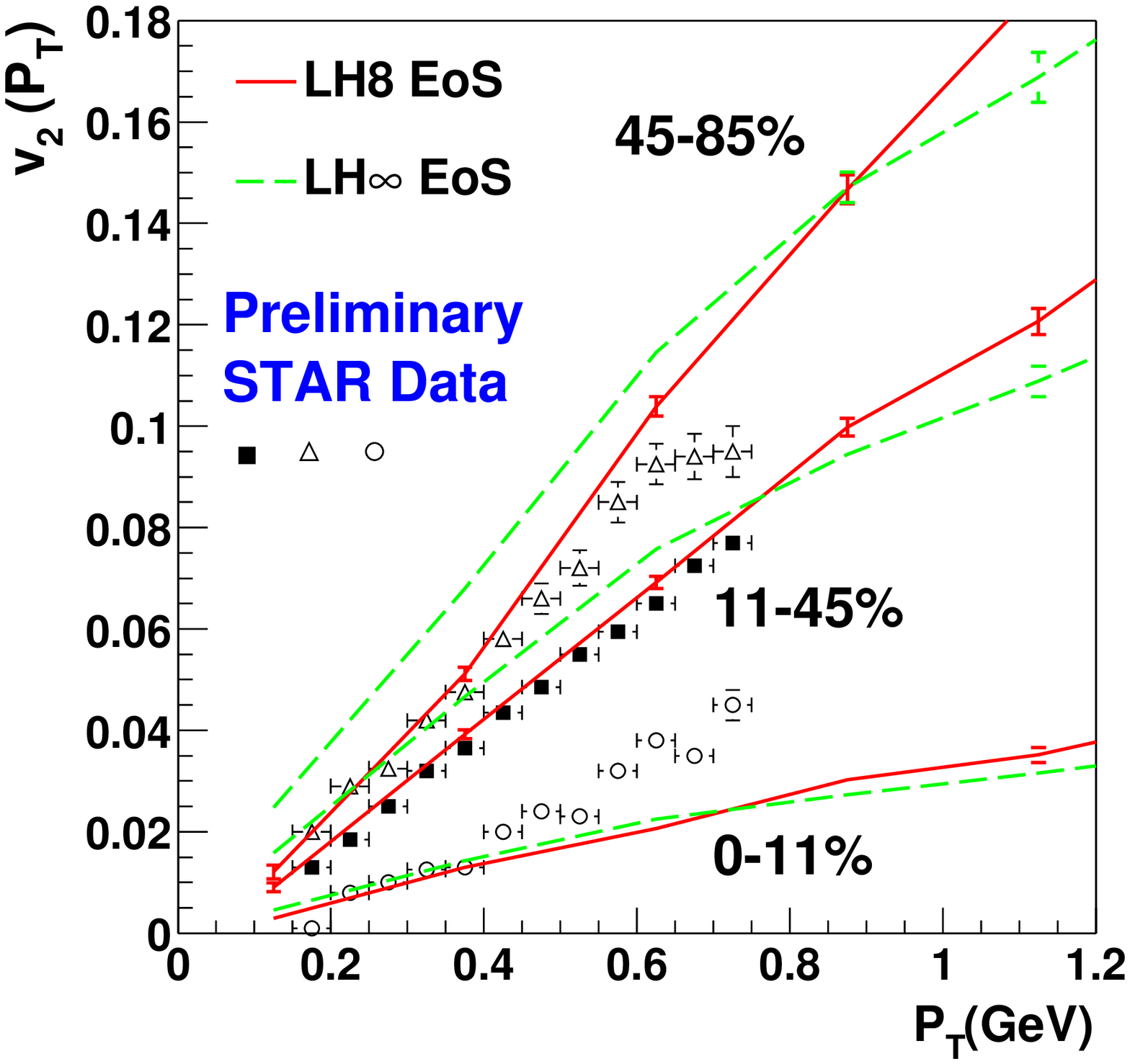}
   \vskip 0.2in
   \caption[]{
      \label{psBscanV2particip}
      v2 versus impact parameter b, described experimentally by the
number of participant nucleons, for RHIC STAR and SPS NA49 experiments. 
 Both are compared to our results, for EoS LH8.
   }
\end{figure}

Let me now jump 
to Fig.\,\ref{psBscanV2particip}(a) which shows the
 impact parameter dependence of elliptic  flow.
First note that at SPS the difference between
theory and data is substantial, especially
at more peripheral collisions. However, this difference completely
disappears\footnote{If STAR data were compared with pure hydro results,
e.g. reported by U.Heinz et al, they are also below. However,
``viscosity''
of hadronic matter at freeze-out, taken care of by RQMD in our model,
naturally explains
 this difference.} at
RHIC,
where even very peripheral collisions demonstrate large elliptic flow.
Another important feature of elliptic flow, also well
 reproduced by hydro
      Fig.\,\ref{psBscanV2particip}(b), is its strong growth with the
transverse momentum.

Last but not least: combining all these hydro results, we were able to
show that the whole body of data (i) cannot be described  by EoS
$without$  the QCD phase transition; and that (ii) it 
is best described by EoS in which the jump in energy density 
(``latent heat'') is about { $\Delta \epsilon\approx 800 MeV/fm^3$}, or  
 $\Delta \epsilon/T_c\approx 8-10$. This happen to be in excellent
agreement
with the lattice results, which the Beilefeld group and others have
predicted
for years.

\section{{ How QGP happened to be produced so quickly at RHIC?} }

Heavy ion physics  entered a new era when Relativistic
Heavy Ion Collider (RHIC) at Brookhaven National Laboratory have
taken
first data in summer 2000. These data
(see e.g. \cite{QM01})
 have shown that
 heavy ions collisions (AA) at highest energies 
significantly differ 
$both$ from the hh collisions 
 and  the AA collisions at lower (SPS/AGS) energies.
Already the very first multiplicity measurements
 by PHOBOS collaboration  
 have shown that particle production per participant
  nucleon is no longer constant, 
as was the case at lower (SPS/AGS) energies, but grows more rapidly.
Long-anticipated semi-hard processes have shown up.

If they are
 perturbative   {\em ``mini-jets''},  cut at scale $p_t> 1.5-2
\,GeV$, the predicted mini-jet multiplicity is expected to be
$dN_g/dy\sim 200$ for central AuAu collisions at $\sqrt{s}=130 \,
AGeV$. 
However,  other  RHIC data
have provided  serious arguments {\em against} the mini-jet  
scenario. Those are:
(i) Strong {\em collective phenomena}
such as  flows desribed above 
 (ii) Jet quenching: Spectra of hadrons at large $p_t$, especially the $\pi^0$
spectra
 from PHENIX,
 agree well with HIJING for peripheral collisions, but  show 
much smaller yields
for central ones.
 Counting from expected Cronin effect
(which in pA collisions is about factor 2)  
we see a suppression  of about order of magnitude. It 
has not been the case at SPS and it can only happen
if the outgoing high-$p_t$
 jets  propagate through dense matter, with $dn_g/dy\sim 1000$.
(iii) Furthermore, this estimate is also supported by STAR data on
elliptic asymmetry parameter $v_2(p_t)$ at large 
transverse momenta $p_t>2\, GeV$, see
 \cite{GW}. Note also that
the result is consistent with
the $maximal$ possible value  evaluated from  the final 
  entropy at freeze-out, $(dN/dy)_\pi\sim 1000$.

In summary, these data are quite consistent with
 the Quark-Gluon Plasma (QGP) (or Little Bang) scenario 
 in which entropy is produced promptly and
subsequent expansion is close to adiabatic expansion of equilibrated
hot medium. 

How this happened remains an outstanding question. One option is
{\em significant reduction of the pQCD cut-off} relative to 
$1.5-2\ GeV$ expected from pp. The other is the subject of the next section.

\section{The instanton/sphaleron mechanism  of entropy
production
at RHIC }

In spite of significant progress related to instanton-induced effects
in QCD
vacuum and hadrons (see e.g. review  \cite{SS_98}),
 very little  has so far been made toward
 understanding  
 high energy processes. This is   mostly because it is difficult 
 to translate many of our non-perturbative tools to
Minkowski space.  

The
  non-perturbative fields in the QCD vacuum (and inside hadrons)
are not some shapeless objects, with
typical size $\sim 1 fm^{-1}\sim 1/\Lambda_{QCD}$, as 
was assumed in the 70's. Instead it is concentrated in $small-size$
instantons, with size $\rho \sim 1/3 fm$, which also generate
``constituent quarks'' of similar size.
When hadrons are boosted to high energies,
they become thin disks: and substructure just mentioned makes their
partons to be
 correlated in the transverse plane. Furthermore,
 what is a part of hadronic wave function in one reference frame, 
becomes the parton-parton interaction  in another. This consideration alone implies
importance of the instanton-induced high energy scattering.

We cannot describe here long history of the so called soft pomeron,
starting from  Pomeranchuck and Gribov in 1960's.
Phenomenologically it is still in very good shape. 
where a supercritical 
pole with the intercept $\Delta\sim 0.08$.
 Perturbative BFKL gluon ladder gives the pQCD version of high energy
behavior, with a supercritical pole with the intercept $\Delta\sim 1/2$: it seem to
work for hard processes, with large $Q$.

 For long time people have constructed multi-peripheral
 models with ladders made of hadrons. Recent story started with 
 Kharzeev and Levin\cite{KL} who kept t-channel gluons but tried to
  substitute the gluonic ``rungs" of the BFKL ladder
by those with a pair of pions, or sigma meson,
 to improve it. They 
used the gg-$\pi\pi$ non-perturbative vertices known
from the low energy theorem. Their estimated value for $\Delta$ 
 was  close to $\Delta_{phen}$.  Introducing $instantons$
into the problem,
 I re-analyzed \cite{Shu_toward} the contribution of the colorless scalar
channel generated by operator $G_{\mu\nu}^2$, using the 
 gg-$\pi\pi$ and gg$-scalar-glueball$ couplings
 determined previously from
the calculation of appropriate Euclidean correlators, see \cite{SS_98}.
 The result turns out to reduce thos of the 
KL paper, with  $\Delta\approx 0.05$ only, and 
  pions and glueball contributions
being roughly equal. 
 
The next important step has been done by two groups,
 Kharzeev, Kovchegov and Levin~\cite{KKL} and M.Nowak,  I. Zahed
and myself~\cite{NSZ}. These works considered inelastic processes, with
multi-gluon production. 
 Basically,
instead of  the glueball peak at $M\sim 1.7 \,GeV$ 
in the
cross section $\sigma_{gg\rightarrow any}(s)$ 
in the {\em colorless scalar} channel, these authors argue that
a particular $colored$ object is produced, 
 the $sphaleron$. The cross section as  a function of the mass
is also believed to have a peak, around 2.5-3 GeV, with a width
given by $classical$ instability of this configuration.
So, in a way, it is a resonance, although not a hadron due to non-zero 
color.

\begin{figure}[ht!]
\begin{center}
  \begin{minipage}[c]{1.6in}
    \centering
     \includegraphics[totalheight=1.6in]{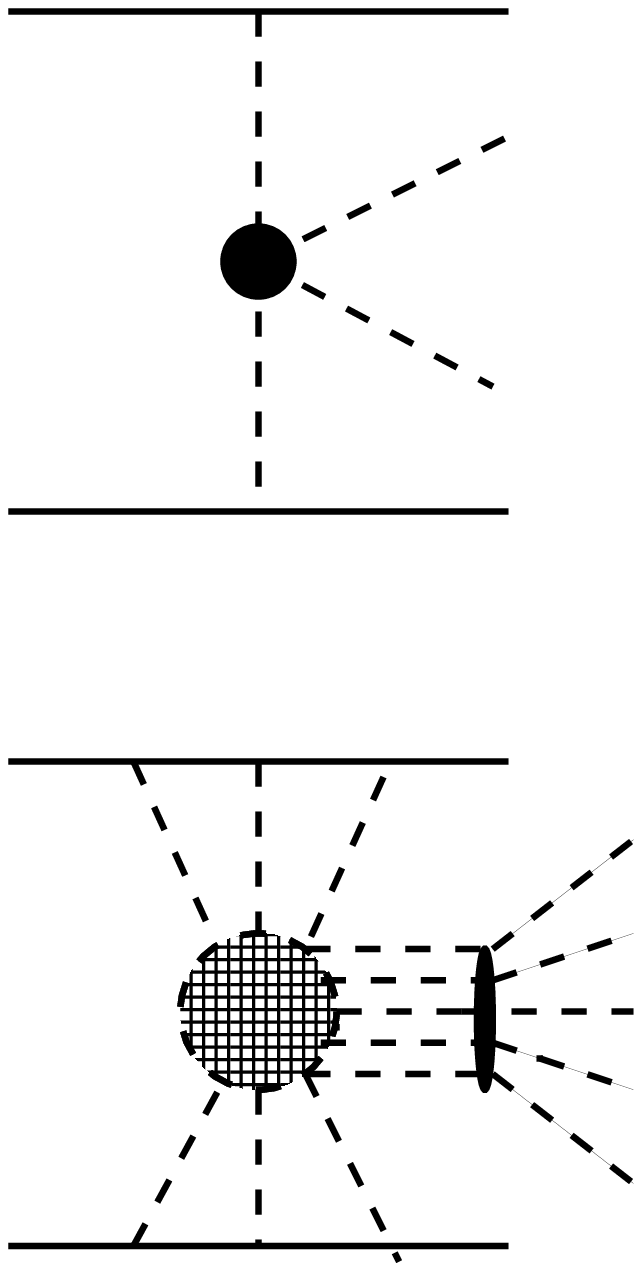}
   \end{minipage}
  \begin{minipage}[c]{2.6in}
    \centering
      \includegraphics[width=2.6in]{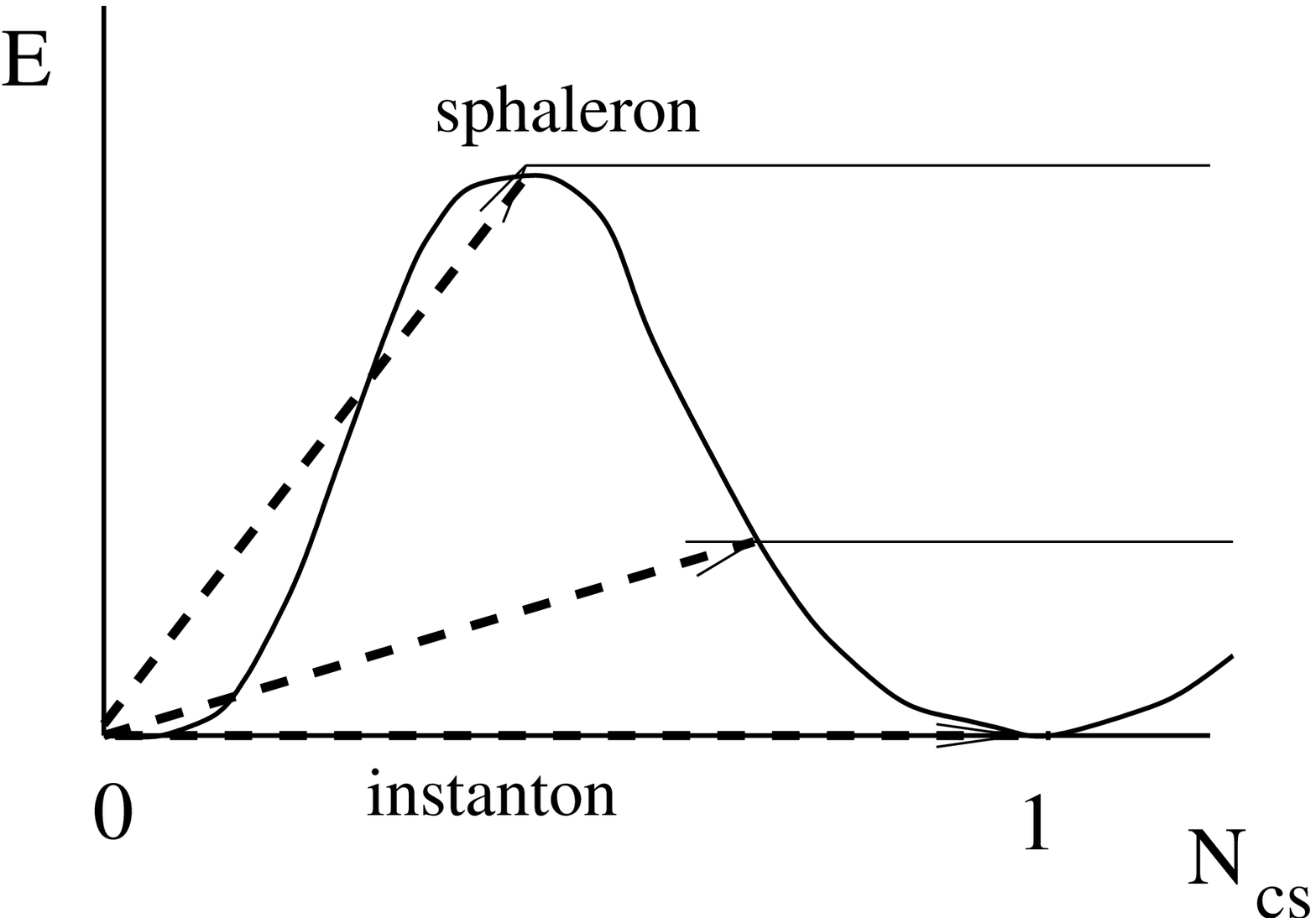}
   \end{minipage}
\end{center}
  \caption[]
  {
   \label{processes}
The lhs of the figure compares
(a) A typical inelastic perturbative process (two t-channel gluons collide,
producing
a pair of gluons) to (b) non-pertubative inelastic process, incorporating
collisions
of few t-channel gluons with the  instanton (the shaded circle),
resulting in multi-gluon production.
 The rhs side of the figure shows the same process in a quantum mechanical way.
The energy of Yang-Mills field versus
the Chern-Simons number $N_{cs}$ is a periodic
function, with zeros at integer points. The $instanton$ (shown by the
lowest
dashed line) is a transition between such points. However if some
nonzero
 energy
is deposited into the process during transition, the virtual path
 (the dashed line) leads to a  {\em turning points},
from which starts the
real time 
motion outside the barrier (shown by horizontal solid lines).
The maximal cross section correponds to the transition to
the top of the barrier, called the
$sphaleron$.
 }
\end{figure}

Hard processes, involving scales $Q^2>>Q_0^2\sim 1 \, GeV^2$
are adequately described by
pQCD: the result of parton collisions is their re-scattering,
see fig.\ref{processes}(a). The multiple production of
N partons is suppressed in pQCD by extra powers of $\alpha_s(Q^2)$.
It is however not true at the $semi-hard$ scale $Q^2 \sim Q_0^2$
because here non-perturbative effects $\sim exp[const/\alpha_s(Q_0^2)]$
become comparable or dominant. Specifically for instantons,
 the so called instanton diluteness parameter\footnote{For comparison,
in electroweak theory it is about $10^{-80}$ or so.} 
 $ (n\rho^4)\sim (1/3)^4\sim 0.01 $,
were n (taken from the phenomenology and/or lattice studies  \cite{SS_98})
is the resulting instanton density in Euclidean space-time, including
all interactions and condensates in the vacuum.
Compared to $(\alpha_s(Q_0)/\pi)^n$ one may expect it to dominate 
over pQCD for processes in the order $n>2$.
And the
non-perturbative phenomena like instanton-induced production indeed
are multi-gluon ones, see fig.\ref{processes}(b).


Note that not necessarily one pair of gluons collides (the figure
depict 3*3 as an example), also several gluons are produced in one act.
This happens  because of the large  field strength
of the instanton $A^{inst}_\mu \sim 
1/g$, which makes processes with any number of gluons of the same
orders in g. 

 Furthermore, the intermediate stage of the
process (shown by
the horizontal dashed lines) indicate $coherence$ of the outgoing gluons:
they are first produce
in the form of specific gluomagnetic field configuration, 
the {\em turning points}, which I study right now \cite{Shu_turning}.

Before we outline results of the
 specific calculations, let me emphasize the basic
quantum mechanics of the process (see the right-hand side of
Fig.1). Instantons are classical solutions
describing
tunneling from one classical vacuum (in which $G_{\mu\nu}=0$
and the $A_\mu$ is pure gauge) to another: naturally the energy is
zero
on it. However if during tunneling some energy is deposited into
classical field, from two or more  colliding gluons, 
after the system appears from under the barrier
it may propagate in real time. Schematic example of that is shown in 
the right hand side of
fig.\ref{processes} for three cases, with deposited energy ranging from
zero (the original instanton) to that of the barrier maximum.

The top point is known as $sphaleron$\footnote{Which means ``ready to
fall''
in Greek.} configuration \cite{Manton}, first
found in the electroweak theory. Intensive studies of the
instanton-induced
processes also were done in this context in early 1990's, driven
basically by possible observability of  baryon number
violating processes in electroweak theory\cite{weakinst}. 
The so called ``holy grail function" showed that processes with
multiple
 quanta production indeed lead to growing cross section, reaching its
maximum
at the sphaleron mass and then decreasing. However,
 since in electroweak theory the maximal cross
section has been found to be still very far from observability,
the interest to this direction have mostly disappeared around 1993 or
so.

The results obtained so far are in reasonable agreement with data,
and explain few qualitative points:
(i)
 The pomeron intercept is  small ($\Delta_{phen}\approx 0.08$ because it is proportional 
 the $first$ power
of the instanton diluteness parameter $(n\rho^4)$.
(ii)
 This mechanism also explains {\em small size} of the Pomeron,
(as seen e.g. from the Pomeron slope $\alpha'(0)\approx 1/(2 \, GeV)^2$):
 the reason is small instanton size. (iii)
 Byproduct: at classical level {\em no odderon} appears. This 
is related to non-trivial property of instantons/sphalerons: they are
always in some SU(2) subgroup of SU(3); and in SU(2) there is no
real distinction between quark and anti-quark.

  In my recent paper \cite{Shu_01} it was suggested that
 the instanton/sphaleron mechanism may be a way toward the 
solution of this ``early entropy'' puzzle.
 In a way, high entropy may come directly
from ``sublimation'' of strong vacuum gluon fields and of the
vacuum quark condensate.

 Assuming it to be the dominant process behind  
 the logarithmic growth of the pp cross section, I
 estimated the 
probability of the sphaleron production directly from data. 
Although the cross section  of prompt production
is {\em surprisingly small} in mean parton-parton
collision, two orders of magnitude below 
  geometric cross section $\pi \rho^2$,
 the total number of parton-parton
collisions in $central$ AA collision is so huge that the 
 number of ``promptly-produced objects''
(mini-jet pairs or sphalerons) in AA
collisions per unit rapidity
is estimated \cite{Shu_01}
${dN_{prompt} \over dy}  \sim 100$.

Although this number is similar to estimated mini-jet production events, this
scenario provides  
significantly larger  amount of the
entropy produced.
Indeed, the mini-jets
 are just plane waves:  
they are classically stable and weakly interacting.
The sphalerons are  kind of
resonances existing already at the classical level.
They  explode into spherical
expanding shells
of strong field, which  rapidly sweep the whole volume
and may convert it into Quark-Gluon Plasma\footnote{In heavy ion case
 partons produced
do not hadronize immediately, as is the case for hh collisions.}
Each QCD ``turning point'' cluster decays into several gluons plus
(with some probablity) up to the whole $\bar u u \bar d d \bar s s$
set
of quarks
 \cite{CS}.
Key signature of these  decays may then be a 
deviation from the ``hot glue scenario'':
 $\bar u u \bar d d \bar s s$ should be there much earlier than it is possible
due to pQCD effects.

\section{Conclusions} 
Let me try to summarize the status of our field, which is 
 a 21-year old youngster, full of life
and experimentation, open to whatever those will bring.
 This
makes this field rather unique among other fields
in high energy and nuclear physics.
Helmut and other ``old-timers'' have all reasons to be proud of it.  
In particular:\\
(i) We finally  have excellent
{ dedicated machine, RHIC}, and look forward for other
observables and for the { LHC era}
for a decade to come. \\
(ii) The field is growing well. It has enough
open questions to provide challenges/
attract many
{\em 1-st rate}  young theorists.\\
(iii) Our understanding of the QCD phase diagram has grown 
enormously. All kind of symmetries -- {\em color, flavor,translations
and even parity} --
can be broken under appropriate conditions.\\  
(iv)
 Amazingly, the Optimists were right: only factor 2 in multiplicity
(from SPS to RHIC) made a lot of difference! 
It is a Bang, not a (so often predicted) fizzle. And yes, we do see
the ``QGP push''.
We seem to finally approach the
macroscopic limit: hydro works, jets are quenched,
ellipticity is very large, even at large $p_t$.
 {\em The EoS is being determined.} Remarkably, it seem
to { confirm $quantitatively$ the values of  $T_c \, and\, \Delta \epsilon$}
what lattice had predicted all along.\\
 Outlook: Yes, there are many questions left open: but
those are being addressed as we speak. One outstanding
question discussed above is {\it how  
so large entropy has been generated so quickly}. 
The other problem I did not mentioned is what exactly is the shape of the freezeout surface, in
view of current HBT data.

\end{document}